# The Evolving Media Discourse on ChatGPT and Higher Education

Yinan Sun
*Information Sciences & Technology*
*George Mason University*
Fairfax, USA
sunyinan@hawaii.edu

Ali Ünlü
*School of Education and Human Development*
*University of Virginia*
Charlottesville, USA
aliunlu@virginia.edu

Aditya Johri
*Informaton Sciences & Technology*
*George Mason University*
Fairfax, USA
johri@gmu.edu

***Abstract*—This research full paper examines how news media have been instrumental in creating specific narratives about generative AI applications, especially ChatGPT, in higher education, and how these narratives have changed over time. The introduction of emerging technologies in higher education is driven not only by their technological affordances but also by the narratives built around their perceived value, risks, and possibilities. Therefore, understanding how news media narratives contribute to sociotechnical imaginaries – the imagined futures of technology use that institutions and educators inherit – is important for evaluating ChatGPT's role in teaching and learning, including engineering education. Through temporal and sentiment analyses of 198 U.S. news articles from November 2022 to October 2024, we traced the evolving narratives surrounding generative AI and the use of ChatGPT in higher education. We found that the media discourse largely centered on institutional responses, with policy changes and teaching practices showing the most consistent presence and positive sentiment over time. Conversely, coverage of topics such as human-centered learning, the job market, and skill development appeared more sporadic, with initially uncertain portrayals gradually shifting toward cautious optimism. Media sentiment toward ChatGPT's role in college admissions remained predominantly negative. Our findings suggest that media narratives prioritize institutional responses to generative AI over the long-term, broader ethical, social, and labor-related implications. This imbalance is especially relevant to engineering and computing education, where students must be prepared not only to use AI tools but also to critically evaluate the broader sociotechnical consequences of AI as future designers of AI-enabled technologies.**

***Keywords— ChatGPT, Generative AI, AI in Education, Sociotechnical Imaginaries, Media Discourse***

## I. Introduction

Artificial intelligence (AI) advancements, including generative AI (GenAI), have created new opportunities for innovation in higher education [1]. A major development occurred with the launch of ChatGPT by OpenAI in November 2023 [2]. Leveraging deep learning, ChatGPT generates human-like responses to a wide range of prompts [3] and supports numerous academic tasks, including personalized learning, writing assistance, grading, content creation, and research support [4]. Its cloud-based accessibility has further expanded its reach across educational contexts worldwide [5].

ChatGPT's rapid integration into higher education has positioned it as a focal point in media discourse, where coverage highlights both its transformative potential and inherent challenges [6-7]. Beyond functionality, media narratives also address broader societal concerns such as ethics, pedagogy, and institutional policy. These narratives can shape sociotechnical imaginaries of AI in education [8], influencing public perceptions, policy development, and future adoption at multiple levels. Rather than a static snapshot, media narratives of GenAI and the sociotechnical imaginaries they construct evolve in response to sociopolitical, institutional, and technological developments. For example, [9] linked peaks in AI news coverage from November 2022 to November 2023 to major tech developments, while [10] observed a broader shift from optimism to caution after ChatGPT's launch.

Although recent studies have explored media portrayals of GenAI [11-12], few have examined how topics and sentiments evolve over time, particularly in the context of higher education. A longitudinal examination is critical for understanding how perceptions of a new technology change. To address this gap, our study investigates how ChatGPT's role in higher education has been discussed in U.S. news media over time. News media in this research refer to both text-based printed and online news formats [13]. We focus on these news outlets due to their enduring influence and credibility, in addition to digital or social media [14]. Print media remain central in shaping public agendas, especially in education and technology discourse. As a leader in AI development, U.S. news sources represent a large extent of the discussion, especially regarding higher education and AI integration [8]. These public narratives are particularly relevant to engineering an computing education because they may influence how faculty and students teach and learn core engineering competencies, such as analytical reasoning, creativity and innovation, sociotechnical problem-solving, ethics, and professionalism. Moreover, engineering students are not only users of generative AI tools but also future designers, developers, evaluators, and decision-makers of AI-driven technologies. We address the following research questions:

1. How have the predominant themes regarding ChatGPT's role in higher education evolved over time in U.S. news media?

2. How have the prevailing sentiments regarding ChatGPT's role in higher education evolved over time in the U.S. news media?

We interpret these evolving themes and sentiments as indicators of sociotechnical imaginaries surrounding ChatGPT in higher education.

Funding information to be added after review.

## II. Related Work

### A. Sociotechnical Imaginaries of GenAI in News Media

The concept of sociotechnical imaginaries refers to the shared vision of desirable futures that societies hold regarding how science and technology will advance social life [15]. In the current landscape, sociotechnical imaginaries related to technology are both prominent and significant, as they influence the regulation and use of GenAI, thereby shaping how it is adopted across domains, such as business, healthcare, and education [8, 16]. News media, such as journalistic practices and editorial choices, play a critical role in constructing and disseminating these imaginaries by framing the risks, benefits, and potential uses of GenAI [17, 18]. In this study, sociotechnical imaginaries provide a theoretical lens for interpreting media narratives as public constructions of what GenAI is, what it is expected to do, which futures it makes possible or problematic in higher education, and how these imagined futures are produced, stabilized, or contested over time.

In [19], authors identify three key sociotechnical imaginaries of GenAI: (1) market–state tensions, where the discourse of individualism and freedom conceals the increasing centralization of power and surveillance; (2) idealized virtual work, which masks the exploitation of free labor behind narratives of innovation and productivity; and (3) the utopia of unlimited knowledge, which celebrates unlimited creation of output while obscuring the lack of understanding or transparency in AI-generated content.

Furthermore, the imaginaries vary across contexts; as [8] observe that UK and U.S. media frequently adopt dystopian framings, emphasizing job displacement and misinformation, in contrast to Chinese and Indian media, which promote utopian visions of AI-driven progress, economic growth, and national leadership in innovation. Although there is a growing body of work on the sociotechnical imaginaries of AI in media narratives, studies that examine news media narratives over time are rare, particularly those related to GenAI in higher education.

### B. Evolving Media Narrative on GenAI

An increasing body of literature has traced the evolution of media narratives concerning the primary topics and sentiments surrounding GenAI, particularly ChatGPT. Regarding the shift in key topics, [9] reported that the subjects of corporate technological development, regulation, and security surged in February and May 2023, uncovering the media's responsiveness to specific social or technological events. [10] noted a thematic shift in media discourse from earlier attention to AI-generated images to a post-ChatGPT emphasis on the quality of AI-generated texts. Similarly, [20] reported that public interest evolved from early enthusiasm to concerns about commercialization, labor, education, and ethics. In terms of sentiments, [21] reported that the emotional reactions of journalists on Twitter/X exhibited more positivity and less negativity after the ChatGPT launch than before.

[22] found that while initial public sentiment toward GenAI tools such as ChatGPT was notably positive, a downward trend in sentiment occurred in January 2023 amid rising concerns about AI's limitations and societal risks. These studies collectively demonstrate a complex and evolving discourse surrounding GenAI, highlighting the pivotal role of news media in shaping public imaginaries of AI. However, how ChatGPT's specific role in higher education has shifted over time remains underexplored.

### C. Topic and Sentiments in Media Narratives of ChatGPT in Higher Education

While GenAI has been discussed across various sectors, news media play a vital role in shaping public understanding of ChatGPT's specific integration into higher education [6, 7, 23]. Despite its importance, there has been minimal academic literature published on the presentation of ChatGPT's role in higher education by news articles. [7] reported that the main themes in news articles from Australia, New Zealand, the United States, and the United Kingdom regarding how ChatGPT affects higher education include concerns about academic integrity, the limitations and weaknesses of AI tool outputs, and opportunities for student engagement in learning.

[23] pointed out that discussions regarding ChatGPT in Malaysian education in newspaper articles highlighted its adoption as the most frequently discussed topic. [6] noted that Malaysian media generally emphasize ChatGPT's potential to transform education, whereas Japanese media focus on the bans against ChatGPT use in schools.

Finally, [24] reported on how news media construct dominant social imaginaries of ChatGPT, highlighting its unaccountable release and disruptive impact on education. These studies provide valuable insight into how news media shape public discourse about the role of ChatGPT in higher education. However, prior studies offer only static snapshots, lacking analysis of how media narratives around ChatGPT have evolved longitudinally.

### D. GenAI, ChatGPT, and Engineering Education

The rapid adoption of GenAI, particularly ChatGPT, is reshaping teaching and learning across higher education, including engineering education [25]. Across higher education, the reception of GenAI has been mixed because its use aligns with some core educational goals while potentially challenging others. On the one hand, it is seen as a valuable resource for creating up-to-date educational materials, assisting with coding and programming practices, supporting virtual simulations, helping design real-world scenarios, and delivering personalized learning experiences. On the other hand, its use raises concerns about ethical implications, overreliance on technology, potential leaks of sensitive information, misinformation and hallucinations, reduced social interaction, and incorrect or biased responses. These tensions are especially consequential in engineering education, where students are expected to develop not only technical competence but also analytical reasoning, sociotechnical awareness, ethical judgment, and professional responsibility [26]. Although prior studies have examined the opportunities and challenges of GenAI in higher education, less attention has been paid to how GenAI is publicly framed by news media and how such framing may shape the understandings of AI literacy, academic integrity, professional preparation, and responsible technology development among engineering educators and students.

## III. Data and Methods

### A. Data Collection and Processing

We collected news articles using Nexis Uni, a leading database frequently used in news content analysis, for articles published between November 2022 and October 2024, coinciding with the initial release of ChatGPT by OpenAI. Using the search terms ("ChatGPT") AND ("higher education" OR "university" OR "college"), we focused on English-language articles that discussed ChatGPT's role in U.S. higher education. The initial search returned 3,313 articles. After removing exact duplicates (same titles) and near-duplicates (same content but different metadata), we narrowed the set to 400 articles. We further excluded articles that lacked a meaningful discussion of ChatGPT in higher education or focused on non-U.S. contexts. The final dataset included 198 articles from 95 news outlets. We collected the article text and metadata, including title, author, date, and publisher.

Initial screening and preprocessing were conducted by the first author using predefined inclusion and exclusion criteria. Dr. Johri contributed to the development of the inclusion and exclusion criteria, the selection of the dataset, and the review of selected articles. Ambiguous cases were discussed collaboratively to improve interpretive consistency and ensure alignment with the study objectives. Because the study aimed to identify thematic and temporal patterns in media discourse rather than test hypotheses, we used computational discourse analysis methods rather than inferential statistical modeling.

### B. Data Profiling

To contextualize our dataset, we categorized the 95 news outlets into four types: university-affiliated, regional/local, national, and business-focused. University-affiliated outlets were the most common (56 sources, 53.7%), followed by regional/local (37, 38.9%), business (4, 4.2%), and national media (3, 3.2%). However, regional and local outlets produced the most articles (89, 44.95%), followed by university-affiliated (56, 28.28%), national (49, 24.75%), and business media (4, 2.02%). This indicates that discourse on ChatGPT in higher education was driven primarily by local and university-based journalism rather than national or business press.

### C. Data Analysis

*1) Examining topic change over time using topic modeling and temporal analysis:* To investigate how media coverage of ChatGPT's role in higher education has evolved, we applied Latent Dirichlet Allocation (LDA) for topic modeling, followed by a two-step temporal analysis involving event extraction and time-series trend analysis. LDA, a generative modeling technique that uncovers latent themes based on word co-occurrence patterns [27, 28], was implemented using the Python Gensim library. LDA was appropriate for addressing R1 because it enabled us to inductively identify recurring themes in news coverage of ChatGPT in higher education without imposing predefined categories. The model was tuned by balancing coherence and perplexity scores across topic ranges from 5 to 10, resulting in a six-topic solution that was deemed optimal [29]. We then extracted AI-related events by using predefined sets of high-probability keywords associated with each LDA topic. Each article was assigned to its corresponding dominant topic. These articles were grouped by month to analyze trends in topic salience over time using Pandas' time-series processing tools.

*2) Examining sentiment change over time using SetFit sentiment analysis:* To assess how sentiment regarding ChatGPT's role in higher education evolved in U.S. news media, we first applied SetFit, is a few-shot text classification approach that fine-tunes sentence-transformer models using a relatively small amount of labeled data [30,31], across 198 articles that totaled 88,960 words and contained 10,058 unique terms. SetFit's contrastive learning paradigm enabled us to train a classifier using the semantic similarity of sentence embeddings generated via the pre-trained all-mpnet-base-v2 model [30, 31]. SetFit is appropriate for addressing R2 because it can capture sentence-level semantics and contextual nuance from a relatively small dataset, making it suitable for analyzing news sentiment toward ChatGPT that may not be adequately detected by simple dictionary-based sentiment tools.

We manually annotated 30 articles for binary sentiment (15 negative, 15 positive), while the remaining 168 articles were inferred through fine-tuned transfer. Training utilized 70% of the labeled data over 5 epochs with a batch size of 4. We evaluated model performance on the remaining 30%, achieving a classification accuracy of 88.89% and an F1-score of 0.886. After assigning sentiment scores using the fine-tuned SetFit model, we calculated monthly averages for each of the six dominant topics, generating normalized scores from -1 to 1 to track sentiment trends over time.

## IV. Results and Findings

To address RQ1, which examined how dominant themes regarding ChatGPT's role in higher education changed over time in U.S. news media, we identified six distinct themes in the topic modeling analysis across the 198 news articles (Table 1). These themes reflect several sociotechnical imaginaries surrounding ChatGPT, including jobs, industry, and young workers (Topic 1), AI/ML skills and career aspirations (Topic 2), collaboration, decision-making, and bias (Topic 3), college admissions and academic integrity (Topic 4), policy, curriculum, and teaching practices (Topic 5), and human-centered learning and teaching (Topic 6).. As detailed in Table 1, the most prevalent discourse centered on Topic 5, accounting for nearly 39% of the dataset, followed by Topic 6 at 21. 21%.

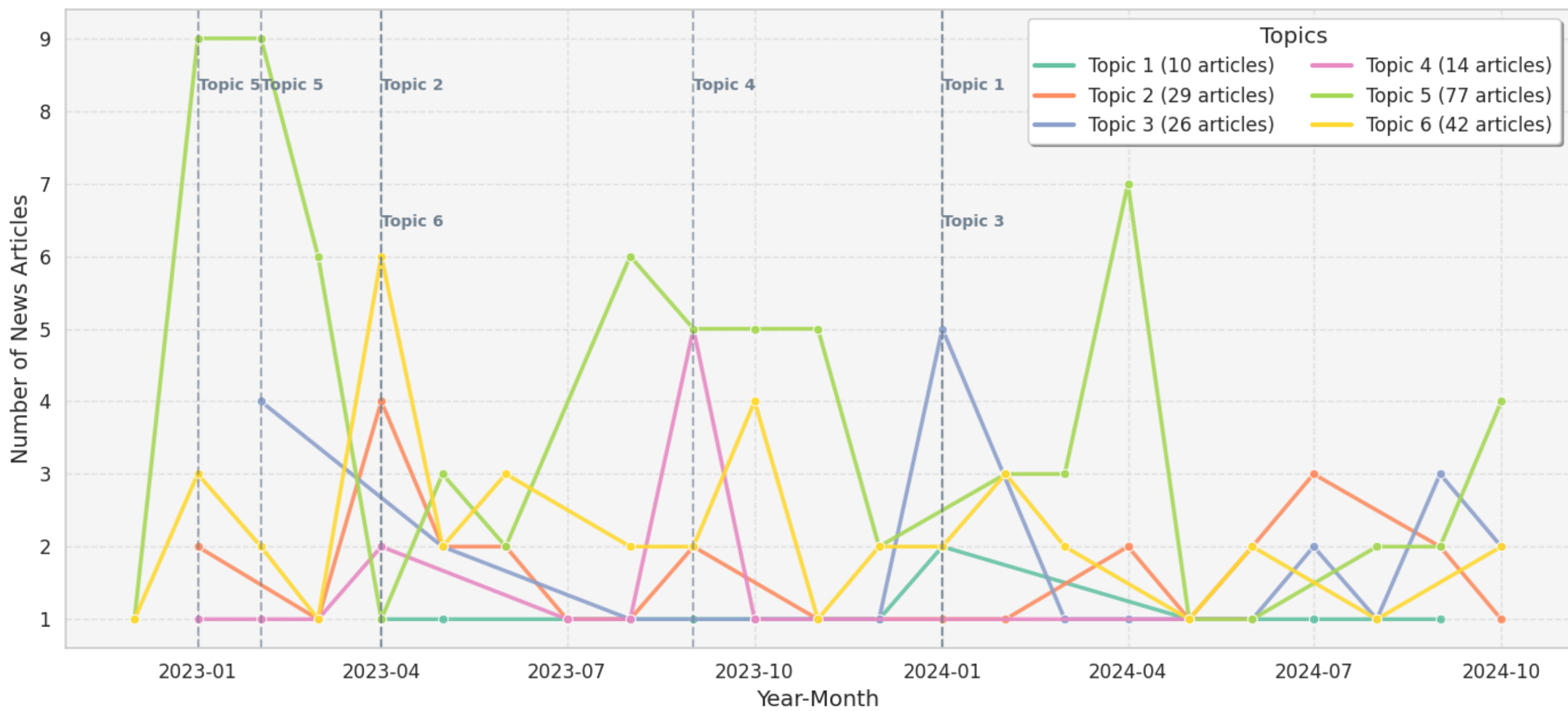

**Figure 1: Temporal trends of news articles by topic.**

## A. *Evolution of Topics Over Time*

The temporal distribution of news articles across six topics shows distinct patterns that reflect evolving interests and focal points in media coverage regarding the role of ChatGPT in higher education (Figure 1). The media narrative began with immediate discussions about how institutions respond to Topic 5, which had the strongest and most sustained presence during the observed period, followed by Topics 6, 2, and 4.

In contrast, Topics 1 and 3 maintained a low presence throughout the period. This suggests that media coverage was more heavily concentrated on topics related to the early stages of adaptation and integration of GenAI tools in higher education, rather than on their impact on post-graduation implications or long-term social issues related to decision-making. Several of these topics are directly relevant to engineering educators and students, including AI/ML skills, decision-making and bias, curriculum and teaching practices, and human-centered learning.

To elaborate on each topic, Topic 1 remained on the margins of media presence throughout the period, indicating that news media coverage of GenAI's long-term impact on the job market is lacking and highlighting a gap in public discourse about this issue compared to education-related topics. For example, the article (no. 63) highlighted growing concerns about mass job displacement and rising inequality driven by GenAI.

Table 1. Major Sociotechnical Imaginaries Identified from Nexis Uni News Articles on ChatGPT in Higher Education.

| | Topic | Description |
|---|---|---|
| 1 | Jobs, Industry, &Young Workers (n=10, 5.05%) | Employment shifts due to ChatGPT, particularly its impacts on young worker. |
| 2 | AI/ML Skills & Career Aspirations (n=29, 14.65%) | Rising AI/ML skill demand and influence on student career goals. |
| 3 | Collaboration, Decision-making, & Bias (n=26, 13.33%) | AI integration in institutions and communities, especially in decision making and healthcare. |
| 4 | College Admissions &Academic Integrity (n=14, 7.1%) | Concerns about ChatGPT in college applications and admission. |
| 5 | Policy, Curriculum, &Teaching Practices (n=77, 38.89%) | Institutional strategies, in response to ChatGPT's classroom integration. |
| 6 | Human-Centered Learning & Teaching (n=42, 21.21%) | Generative AI's role in personalized, human-centered education. |

Topic 2, while receiving limited early attention, maintained relatively steady media visibility across the study period. A significant peak was observed in April 2023, following a declining trend in 2024. Coverage during peak time addressed both the expansion of AI-focused academic programs and growing uncertainties about the future of tech-related careers. For instance, the article (no. 55) highlights the University of Wisconsin–Madison's $260 million investment in a new Computer, Data, and Information Sciences (CDIS) building to support booming tech majors. Another article (no. 53) narrates a personal account of a student switching from computer science to finance due to concerns that AI would soon surpass human programmers.

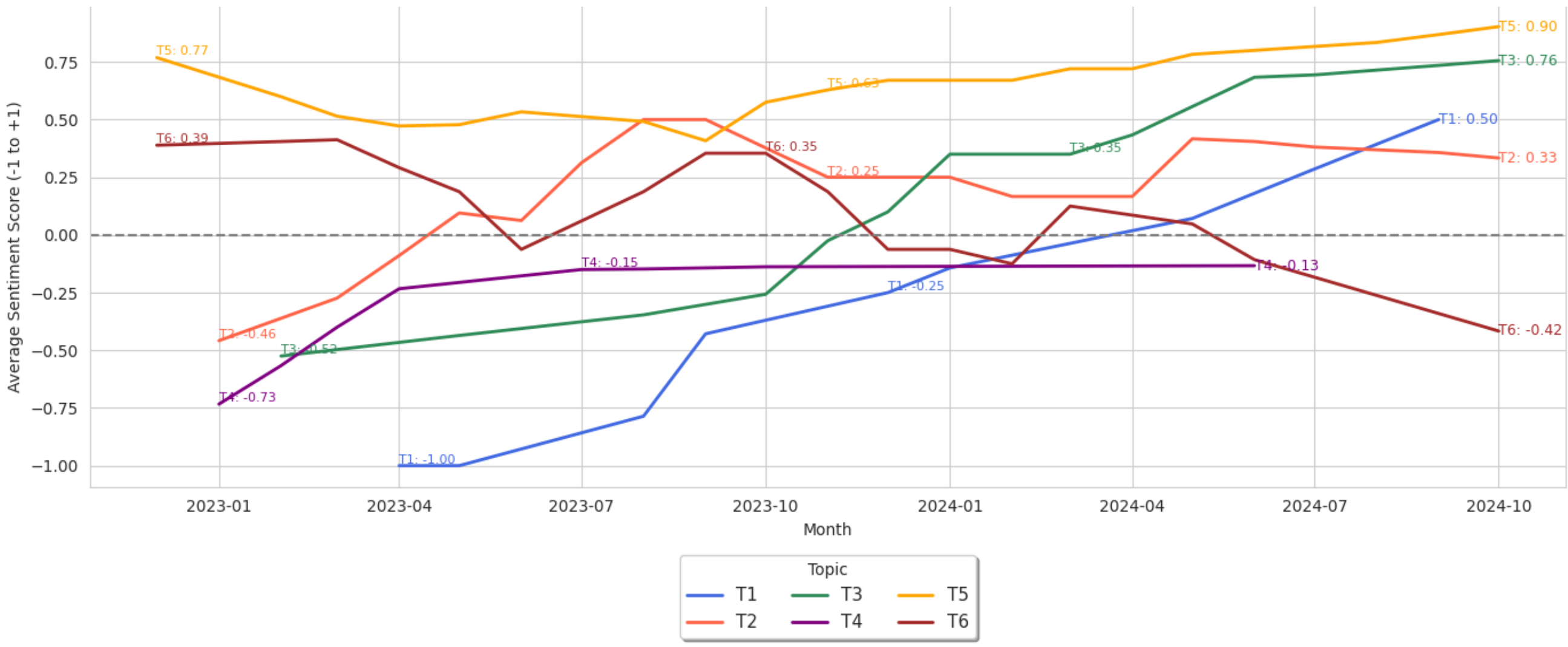

**Figure 2: Monthly sentiment trend by topic.**

Topic 3 garnered moderate media attention in early 2023 and received renewed focus in mid-to-late 2024. While it did not emerge as a dominant theme, this trend indicates a gradual shift in media interest toward exploring the deeper, long-term societal and ethical implications of GenAI. A notable peak occurred in January 2024, coinciding with OpenAI's announcement of ChatGPT Team, a service designed to support collaborative workspaces and provide administrative tools for team management. This surge also paralleled academic and institutional developments. For instance, a study led by Stanford researchers revealed that widely used AI chatbots may reinforce racial misconceptions in healthcare (no. 129). Additionally, Arizona State University became the first university to partner with OpenAI (no. 135).

Topic 4 attracted minimal media attention in the early stages but received a brief, concentrated focus in September 2023. Following this short surge, coverage diminished significantly, suggesting a more episodic media interest related to the college admissions cycle rather than ongoing discussion by the news media. The articles (no. 88; no. 89) explore how AI tools like ChatGPT reshape college admissions, particularly the role of personal essays.

Topic 5 received immediate and intense media attention at the start of 2023 and maintained a steady presence in media coverage throughout the observed period. This early spike in coverage coincided with OpenAI's release of a complementary tool aimed at helping teachers detect plagiarism, exemplified by the article (no. 17), which reflects the initial wave of reporting that explored how educational institutions were responding to ChatGPT's classroom integration. Although coverage declined somewhat after this early peak, it rose again in August 2023. This period reignited discussions around AI integration into teaching and learning (no. 12).

A third noticeable spike occurred in April 2024, aligning with summer school preparations and broader pedagogical planning cycles. For example, the article (no. 152) reported that school board members debated whether discussions of ethics and bias should be part of the curriculum for a proposed AI summer program led by Stanford and MIT graduates. These recurring spikes suggest that media discourse around Topic 5 has remained consistently active and responsive, often aligning with major announcements from technology companies or key moments in the academic calendar.

Topic 6 shows a moderate yet consistent media presence throughout the timeline, with regular mentions in 2023 and 2024. This suggests a sustained interest from news media in how AI tools can impact personalized, human-centered education, despite an overall declining trend. The first major spike occurred in April 2023, shortly after the launch of ChatGPT-4 in March. One illustrative article (no. 48), published in April 2023, documents how professors respond to the growing presence of AI tools in the classroom.

### B. *Evolution of Sentiments Over Time*

To address RQ2, we examined how the prevailing sentiment associated with each topic changed over time. The sentiment distribution across the six topics reveals distinct patterns regarding ChatGPT's evolving role in higher education discourse (Figure 2). Overall, Topics 1, 2, 3, and 4 initially exhibited predominantly negative sentiment in news media coverage. However, Topics 1, 2, and 3 gradually shifted toward more positive sentiment over time, while Topic 4 maintained a consistently negative trajectory, reflecting ongoing concerns about employing ChatGPT for college applications. In contrast, Topics 5 and 6 began with positive sentiment. Topic 5 continued to gain increasingly favorable sentiment over time, whereas Topic 6 experienced a downward shift, ending with a more negative sentiment, highlighting growing apprehension around the limitations of GenAI in supporting human-centered education.

Topic 1 initially exhibited the most negative sentiment in early 2023, reflecting the initial anxiety about GenAI's potential to displace jobs and disrupt career pathways for young professionals. However, sentiment steadily improved over the following months, turning positive by mid-2024 and ultimately reaching an even more positive state by October 2024. Representative articles illustrate this sentiment shift, ranging from highly negative coverage in May 2023 (no. 63) to more optimistic portrayals by July 2024 (no. 173). This upward trajectory suggests a shifting narrative in which fears of automation were gradually replaced by optimism regarding new career opportunities, workforce innovation, and the evolving demand for AI-related competencies.

Topic 2 began with a moderately negative sentiment in early 2023, reflecting initial skepticism and concern over the impact of GenAI on career prospects. However, sentiment steadily improved, reaching a peak in July and August 2023, before declining slightly and ultimately stabilizing at a positive level by late 2024. For example, the article (no. 53), published in April 2023, documents an individual student's decision to switch from computer science to finance due to concerns about AI automation. Sentiment gradually increased, as illustrated by Brigham Young University's announcement of a new machine learning major beginning in fall 2023 (no. 74). This trajectory is complicated by the article (no. 172), which highlights growing uncertainty among students about the long-term value of their degrees amid AI's influence on the job market. Overall, the shift in sentiment reflects a transition from skepticism to a more strategic view of AI/ML skills as essential for future careers.

Topic 3 began with negative sentiment in early 2023, likely due to concerns over AI's potential to perpetuate bias, misinformation, and opaque decision-making processes. However, the sentiment showed consistent improvement over time, turning positive in late 2023 and peaking by October 2024. Representative articles illustrate this shift. For example, the article (no. 30) exemplifies early-stage criticism, reporting backlash after the university used ChatGPT to generate a condolence message. In contrast, the article (no.192) describes how the City University of New York secured a $200,000 grant to develop ethical AI teaching tools. This progression indicates a growing trust in the use of AI for enhancing collaborative learning and augmenting the decision-making process.

Topic 4 maintained a consistently negative sentiment, showing only marginal improvement over time but remaining below the neutral threshold. It reflects ongoing concerns regarding plagiarism, dishonesty, and the ethical implications of using GenAI in admissions processes. For example, the article (no. 19) mentioned that Emory faculty and students express divided views: some embrace ChatGPT, while others highlight integrity concerns. While this article (no. 19) presents a more balanced tone, most articles published in September (e.g., no. 102; no. 90) raised practical and ethical concerns over authenticity and fairness in the college application process. This persistent negativity suggests that media discourse often frames the use of GenAI in the college admission process with a more cautious and critical perspective.

Topic 5 consistently exhibited the most positive sentiment among all topics and became increasingly favorable by October 2024. For example, the article (no. 14) illustrates early optimism, published in early 2023. By the end of 2024, articles (e.g., no. 186; no. 195) reflect growing institutional investment in embedding AI into teaching and learning environments. This trajectory reflects sustained enthusiasm for AI's role in reshaping higher education regarding policy and practice.

Topic 6 started with a moderately positive sentiment in early 2023. However, unlike other topics, its sentiment steadily declined over time, turning negative by mid-2023 and reaching an even more negative level by October. This reversal indicates a growing unease about AI's role in potentially disrupting or even replacing traditional learning experiences or constraining the agency of educators and students. To be specific, early media coverage (e.g., no. 11) highlighted efforts to integrate AI ethically and creatively in classrooms, but later articles (e.g., no. 73; no. 179) emphasized transparency issues, ethics, additional labor burden for educators, and fears that AI may undermine students' critical thinking.

## V. Discussion

The patterns identified through temporal and sentiment analysis offer important insights into how U.S. news media construct sociotechnical imaginaries of ChatGPT in higher education. Firstly, the temporal patterns reveal a framing process that prioritizes institutional adaptation, such as curriculum, policy, and practice changes, while underrepresenting other long-term societal, ethical, and economic concerns. Secondly, the main sentiments were initially cautious, particularly regarding academic integrity and labor implications, but gradually shifted toward optimism, highlighting new job opportunities, AI/ML skill development, and improved collaborative decision-making.

Regarding the temporal distribution, our findings highlight a critical imbalance in how U.S. news media have framed ChatGPT in higher education. Topic 5 received immediate and persistent media attention throughout the observed period, suggesting a media narrative that frames GenAI as a tool for institutional responsiveness and pedagogical enhancement rather than a disruptive threat.

Other topics, such as Topics 2, 3, and 6, received less but moderate media attention. This may be because these issues are still related to the early stages of ChatGPT adoption. Notably, these discussions are framed primarily at the individual level (e.g., faculty, students) rather than on the institutional level. This framing highlights how media coverage can direct the public's attention toward certain aspects of GenAI in higher education and influence how these issues should be perceived.

Topic 1 remained largely absent from mainstream coverage, despite being a key focus in academic discussions on the societal implications of AI and employment (e.g., [32]). This divergence also reflects a long-standing critique in media studies that news outlets often overlook labor and equity issues when framing technological innovation [33].

Our findings also demonstrate a notable divergence between academic discourse and media narratives for each topic. Despite a growing body of research on ChatGPT's implications for college applications and admissions, Topic 4 has appeared in the news only sporadically. Topic 2 (AI/ML Skills and Career Aspirations) and Topic 6 (Human-Centered Learning and Teaching) are also emphasized in scholarly literature as critical areas of transformation in the age of AI (e.g., [34]). However, media attention to these issues has been moderate and has decreased over time, indicating differing priorities between academic concerns and media framing (e.g., [6]).

The spikes for each topic tend to align with major technological developments. Specifically, our findings show that Topics 2, 3, 5, and 6 all reached their peaks right after the launch of new versions or features of ChatGPT. Future research should further investigate how major technology companies can potentially influence the public's perception through media narratives.

Regarding the evolution of sentiment, our findings indicate that U.S. news media have shifted from a negative sentiment when framing the role of ChatGPT in higher education, with an increasing positivity over time. This trend contrasts with earlier studies that observed a general shift in media sentiment from initial optimism to heightened caution and skepticism following the launch of ChatGPT (e.g., [10, 21]). One explanation for this divergence lies in the domain-specific framing practices: whereas coverage in sectors such as security, economics, or health often emphasizes risk and bias, media narratives on higher education tend to focus on innovation, policy responsiveness, and pedagogical transformation [10]. Another explanation can be an increased understanding of ChatGPT regarding its capabilities, which resulted in reduced skepticism and improved adoption. Furthermore, our findings provide a more nuanced perspective by examining how sentiment varies across each topic over time, capturing phases of initial skepticism, excitement, and adaptation toward technology.

Although the overall positive sentiment pattern aligns with the temporal pattern, our analysis reveals the limits of media optimism: while supporting institutional efforts to integrate AI into policy and pedagogy, skill development, and decision-making, there is a resistance to its use in matters of fairness concerning college applications and admissions, as well as in replicating human agency in education. These findings provide a more nuanced understanding than prior studies [9-10] that suggest news media, particularly in the Global North, tend to adopt a more critical or alarmist tone when reporting on ChatGPT and generative AI. Specifically, Topics 1, 2, and 3 were initially framed negatively but shifted to a more positive frame over time. In contrast, Topic 4 was continually portrayed as negative throughout the observed period. Topic 6 features an initially positive yet declining trend over time. This framing illustrates the multilayered sociotechnical imaginary of GenAI: the potential for institutional transformation, creativity, workforce innovation, and possible disruptions to academic integrity and erosion of human agency, echoed in social media discourse [35] and academic literature [36].

We end our discussion by revisiting the central concept that frames our work, sociotechnical imaginaries. As the topical analysis indicates, the key sociotechnical imaginaries of AI identified by [19] are represented in some form within our data. Primarily, the optimistic aspects are more prominent, whereas the criticisms are lacking. The discourse is about what is possible and what can be done, i.e., individualism and freedom, but the discussion of the ever-present danger of losing power and privacy to technology companies is lacking. There is also very little emphasis on how these systems are created through free labor and copyright violations. Interestingly, the AI literacy that will be needed to adopt and use these systems productively is also not often mentioned. A key feature of the American sociotechnical imagination, as identified by [37], is that technology's benefits are seen as unbounded whereas risks are usually framed as limited in scope and manageable. Our analysis of the U.S. news media mirrors their observation. The framing of topics was positive and risks, when discussed, were construed as being something that could be handled, especially as the technology developed and became more robust. A final observation we make here is about the "elitism" of the discourse [38], both in terms of who covers the topics, who gets the coverage, and how much of the discourse is driven by the technology companies themselves. We hope this analysis and the discussion led to a more nuanced understanding of technology, in particular, the emphasis on the lack of neutrality, and contribute to a deeper understanding of how knowledge about GenAI is being produced through media.

In engineering education, our findings suggest that educators should critically engage with media narratives as pedagogical resources to examine students' expectations regarding AI use, academic integrity, professional preparation, and responsible technology development. For example, instructors can use media narratives as discussion materials to help students evaluate AI's promises and risks, connect institutional adaptation with broader sociotechnical consequences, and reflect on their future responsibilities as designers and decision-makers of AI-enabled systems.

Several limitations of this work highlight avenues for future research. First, our dataset was confined to Nexis Uni and a specific timeframe, with most articles coming from local and regional news outlets; therefore, the findings may not fully represent national media discourse. However, local and regional reporting also plays an important role in shaping institutional and public understandings of emerging technologies within higher education contexts. Future work could use broader sources and extended timelines to reveal more nuanced trends and compare regional and national media narratives to examine potential differences in sociotechnical imaginaries. Second, focusing on U.S. media excludes global perspectives, and given the reach of news media in the current global world, it can be assumed that non-U.S. sources also play a role in shaping the discourse. Third, because this study examines broader U.S. news media discourse on ChatGPT in higher education, the findings should not be interpreted as representing engineering education discourse specifically. Future research should examine engineering-focused media coverage to better understand how GenAI imaginaries are translated into engineering education contexts. Fourth, this study focused specifically on written-form media discourse, excluding images and headlines as additional interpretive dimensions. Future studies could incorporate multimodal media analysis to examine how visual framing and headlines shape sociotechnical imaginaries of GenAI in higher education. Fifth, media reporting may contain editorial and institutional biases. To reduce the influence of any single source or outlet type, we included articles from multiple news outlets within Nexis Uni, applied consistent inclusion and exclusion criteria, and interpreted the findings as part of how sociotechnical imaginaries are constructed and communicated to the public rather than objective assessments of ChatGPT's impact. Sixth, LDA may oversimplify articles with overlapping themes—future work should explore more flexible models such as Top2Vec or GPT-based clustering. Seventh, our sentiment analysis utilized a small dataset and binary classification; larger datasets and multi-class models (e.g., RoBERTa, DistilBERT) could provide more detailed insights. Finally, comparing framing across academic, social, and news media would further clarify how narratives around GenAI are shaped across platforms.

## VI. Conclusion

This study examined how U.S. news media narratives about ChatGPT in higher education evolved from November 2022 to October 2024. Our findings highlight a focus on early institutional and pedagogical concerns over broader social, ethical, and economic implications, accompanied by a general movement from initial caution to a more positive sentiment across most topics. These insights enhance our understanding of how media shapes sociotechnical imaginaries of AI in education and provide guidance for educators, policymakers, and technologists aiming to promote equitable and responsible AI integration. For engineering education, this study provides a critical lens to help educators and students engage with GenAI beyond its immediate instructional uses by examining issues that are less visible in news media but highly relevant to engineering education.

## VII. ACKNOWELDGEMENTS

This work is partly supported by U.S. NSF Award# 2439459, 2439460, 2319137, 1954556, and USDA/NIFA Award# 2021-67021-35329. Any opinions, findings, and conclusions or recommendations expressed in this material are those of the authors and do not necessarily reflect the views of the funding agencies.